\documentclass[aps,prd,nofootinbib,superscriptaddress]{revtex4}
\usepackage[sumlimits]{amsmath}
\usepackage{epsfig}
\usepackage{amssymb}
\usepackage{graphicx}
\usepackage{pstricks}
\usepackage{color}
\usepackage{bbold}
\usepackage{hyperref}

\begin{document}
%
%
\title{Emission of spacetime waves from the partial collapse of a compact object}

\author{Emmanuel Alejandro Avila-Vargas}
\email{emmanuel.avila5664@alumnos.udg.mx}
\author{Claudia Moreno} 
\email{claudia.moreno@academico.udg.mx}
\author{Rafael Hern\'{a}ndez-Jim\'{e}nez}
\email{rafael.hjimenez@academicos.udg.mx}
\affiliation{Departamento de F\'{i}sica, CUCEI, Universidad de Guadalajara \\ Blvd. Marcelino García Barragán 1421, C.P. 44430, Guadalajara, Jalisco, M\'{e}xico}

%
%
\date{\today}

\bigskip
\begin{abstract}
In this work we describe the partial collapse of a compact object and the emission of spacetime waves as a result of back-reaction effects. As a source mass term we propose a non-smooth continuous function that describes a mass-loss, and we then obtain the solution of such setting. We present three distinct examples of the evolution of the norm $|R_{nl}(t,r_{*})|$ in terms of $t$, and four different results are shown for the parameter $l=1,2,5,10$; here $r_{*}$ is the fixed radius of an observer outside the compact object. In all cases, the decay behaviour is actually present at $t\gg 1$ and becomes more evident for larger $l$. In addition, for the results that have smaller $l$'s their amplitudes are larger when the asymptotic character of $|R_{nl}(t,r_{*})|$ clearly appears. Finally, the farther away an observer is set, the fewer oscillations are perceived; however, from our particular fixed set of parameters, the best spot to observe the wiggles of the emitted spacetime waves is close to $r_{*}\simeq \alpha$. 
\end{abstract}
\keywords{gravitational waves}

\maketitle

%
%
\section{Introduction}
The General Relativity Theory (GR) \cite{Einstein:1915ca} created by Albert Einstein has become the most compelling theory of gravitation, since this framework can successfully explain almost all known gravitational phenomena. We can name the small perihelion of Mercury \cite{Einstein:1915bz}, the immeasurable cosmology physics \cite{Baumann:2022mni}, the shadow of the black hole Sagittarius A*
\cite{EventHorizonTelescope:2022xnr}, to the most recent discoveries of gravitational waves (GW) due to the collapse of binary black holes (BBH) and binary neutron stars (BNS)~ \cite{Abbott_2019, abbott2020gwtc2}.

To explain the dynamics of GW one implements the linearized Einstein equations, which is a perturbative approach, and, in fact, many astronomical events such as the collision of massive black holes, supernovas, and; even the early expansion of the universe could be considered as a source of primordial GW \cite{Baumann:2007zm}. However, a novel non-perturbative method was introduced by considering the Lagrangian formulation of GR from the Einstein-Hilbert (EH) action, where the underlying spacetime manifold has a boundary $\partial V$, that generates a back-reaction effect, which in turn produces spacetime waves (SW)~\cite{Ridao:2015oba, Ridao:2014kaa,HERNANDEZ2020100424}. This new scheme allows us to describe the propagation of such SW in a more general way; for instance, we can use different background metrics and not only a Minkowski's one. In fact, in this work we will carry out our calculations within a semi-Riemannian manifold to describe boundary conditions~\cite{Ridao:2015oba}.

On the other hand, the collapse of a compact object is a growing topic of study from the astrophysical to the theoretical points of view. For example, investigations of a spherically symmetric collapse driven by a scalar field~\cite{2000gr.qc.....1046G, 2004PhRvD..69b7502G, 2005CQGra..22.2295G}; and more recently, this issue has been explored from back-reaction effects during a partial time-dependent collapse of a spherically symmetric compact object~\cite{HERNANDEZ2020100424}, where authors utilised a smooth continuous mass source term. However, in this work, we will study the same phenomenon but now having as a source mass term a non-smooth continuous function that describes a mass-loss $M=M(t)$, in which the initial mass is $M_0$ and the final mass is $M_0/2$. Moreover, recall that GW are produced by symmetry breaking, so we consider a spherical non-rotating compact object, which loses mass, here is the rupture of symmetry, as an attempt to relate SW to GW.

This paper is organized as follows. In Section~\ref{GWs_formalism} we introduce a new SW formalism. In Section~\ref{toy_model}, we study SW's production due to the partial collapse of a compact object. Finally, in Section~\ref{conclusions} we will give our final discussions.

%
\section{Waves of spacetime due to back-reaction effects}\label{GWs_formalism}
We can get Einstein's Field Equations (EFE) using physical principles like the conservation of energy-momentum tensor, the Newtonian gravity limit; this approximation states that when we have a weak gravitational field we get Poisson's equation for the gravitational potential. In addition, we need the weak and strong equivalence principle to establish the relation between physics and geometry \cite{Wald:1984rg}. However, due to EFE preservation under certain kind of transformations, this is not the only way of deriving them. As a consequence of the aforementioned, a more mathematical way of getting EFE was developed, using variational principles and the Einstein-Hilbert action ($I_{EH}$), defined as follows:
\begin{equation}\label{eq:1}
I_{EH} = \int_{V}d^{4}x\frac{\sqrt{-g}}{2\kappa}\left(R + 2\kappa\mathcal{L}_{M}\right) \,,
\end{equation}
where $\kappa = 8\pi G$~{\footnote{In the entire paper we use natural units: $c=\hbar=1$}}, $V$ is the volume of the four-dimensional manifold that represents the spacetime; therefore, this underlying geometry has a boundary $\partial V$. Furthermore, $g$ is the determinant of the covariant background tensor metric $g_{\mu\nu}$, $R=g^{\mu\nu}R_{\mu\nu}$ and $R^{\alpha}_{\,\mu\nu\alpha}=R_{\mu\nu}$ are the scalar curvature and the Ricci curvature tensor, respectively. They are derived from the curvature tensor $R^{\alpha}_{\,\beta\gamma\delta}= \Gamma^{\alpha}_{\,\beta\delta\,,\gamma} - \Gamma^{\alpha}_{\,\beta\gamma\,,\delta} + \Gamma^{\epsilon}_{\,\beta\delta}\Gamma^{\alpha}_{\,\epsilon\gamma} - \Gamma^{\epsilon}_{\,\beta\gamma}\Gamma^{\alpha}_{\,\epsilon\delta}$, where Christoffel symbols are written in terms of the metric tensor and its partial derivatives $\Gamma^{\sigma}_{\alpha\beta} = \left(g_{\gamma\beta\,,\alpha}+g_{\gamma\alpha\,,\beta}-g_{\alpha\beta\,,\gamma}\right)g^{\sigma\gamma}/2$. The Greek indices run from 0 to 3, additionally if latin indices $m$, $n$, etc. appear, they go from 1 to 3. Then, $\mathcal{L}_{M}$ is the matter field Lagrangian, and $V$ is the volume of the four-dimensional manifold that represents the spacetime. We then impose the condition of a stationary action ($\delta I_{EH} = 0$), which yields:
\begin{equation}\label{eq:8to12}
\delta I_{EH} = \frac{1}{2\kappa}\int_{V}d^4x \sqrt{-g}\left[\delta g^{\mu\nu}\left(R_{\mu\nu}-\frac{1}{2}g_{\mu\nu}R -\kappa T_{\mu\nu}\right) + \underline{g^{\mu\nu}\delta R_{\mu\nu}} \right] = 0 \,,
\end{equation}
where $T_{\mu\nu}$ is the stress-energy tensor defined by:
\begin{equation}\label{stress-energy_tensor1}
T_{\mu\nu} = g_{\mu\nu}\mathcal{L}_{M} - 2\frac{\delta\mathcal{L}_{M}}{\delta g^{\mu\nu}}  \,.  
\end{equation}
Note that the expression in parentheses in eq. \eqref{eq:8to12} is the usual EFE; and we will examine with great detail the underlined term in the next section, which will be relevant due to the boundary contribution of the spacetime.  
\subsection{Treatment of the boundary term}
The term $g^{\mu\nu}\delta R_{\mu\nu}$ should not contribute to the field equations, since it contains second derivatives of the metric tensor; therefore, the dynamic equations become of order higher than two. Moreover, note that the variation $\delta I_{EH} $ is integrated with respect to the natural volume element of the covariant divergence of a vector; then we can apply Stokes' theorem and, consequently, evaluate $g^{\mu\nu}\delta R_{\mu\nu}$ at the boundary $\partial V$. In fact, historically Hawking-Gibbons-York (HGY) proposed adding a counterterm to the EH action, which relates the boundary constraint and extrinsic curvature \cite{Gibbons:1976ue, York:1972sj}, to eliminate the contributions coming from $g^{\alpha\beta}\delta R_{\alpha\beta}$. However, if there were any relevant physical phenomena, they are immediately erased. In this work, we will explore an alternative proposal, a new scheme in which the boundary term is considered as a flux of SW over a hyper-surface $\partial V$~\cite{Ridao:2015oba, Ridao:2014kaa}. First, note that 
\begin{equation}\label{eq:19}
g^{\mu\nu}\delta R_{\mu\nu} = \nabla_{\sigma}\left(g^{\mu\nu}\delta\Gamma^{\sigma}_{\mu\nu}-g^{\mu\sigma}\delta\Gamma^{\nu}_{\nu\mu}\right) \,, 
\end{equation}
where $\delta\Gamma^{\sigma}_{\mu\nu}$ is an arbitrary variation of the connection, introduced by replacing $\Gamma^{\sigma}_{\mu\nu}\rightarrow \Gamma^{\sigma}_{\mu\nu}+\delta\Gamma^{\sigma}_{\mu\nu}$. We can associate this term with a 4-vector $\delta W^{\sigma} = g^{\mu\nu}\delta\Gamma^{\sigma}_{\mu\nu}-g^{\mu\sigma}\delta\Gamma^{\nu}_{\nu\mu}$, so $\nabla_{\sigma}\delta W^{\sigma}$ is the 4-divergence of this vector field (due to Stoke's theorem, as we stated before). Therefore, we take the covariant derivative and make a contraction of the indexes, yielding: 
\begin{equation}\label{eq:19a}
g^{\mu\nu}\delta R_{\mu\nu} = \delta W^{\sigma}_{\,;\sigma} \,, 
\end{equation}
where $\delta W^{\sigma}_{\,;\sigma}=\nabla_{\sigma}\delta W^{\sigma}$ is the short notation of the covariant derivative. Second, given that the result of the divergence of this tetra-vector $\delta W^{\mu}$ produces a scalar flux, whose origin is purely geometric, we identify this outcome in the following way:
\begin{equation}\label{eq:25} 
g^{\mu\nu}\delta R_{\mu\nu} = \delta W^{\sigma}_{\,;\sigma} = \delta \Phi(x^{\alpha}) \,,
\end{equation}
and this geometric scalar field $\delta \Phi=\delta \Phi(x^{\alpha})$ can be related with the cosmological constant $\Lambda$ by:
\begin{equation}\label{eq:26}
\delta \Phi = \Lambda g^{\mu\nu}\delta g_{\mu\nu} = -\Lambda \delta g^{\mu\nu} g_{\mu\nu} \,.
\end{equation}
Hence, having the condition $\delta I_{EH} = 0$ and given that the variation $\delta g^{\mu\nu}$ is arbitrary, we shall obtain: 
\begin{equation}\label{eq:30}
R_{\mu\nu}-g_{\mu\nu}\left(\frac{1}{2}R+\Lambda\right) = \kappa T_{\mu\nu} \,.
\end{equation}
We have, in fact, got the EFE with a cosmological constant, which is connected with the boundary flux. The above scheme presents a distinct derivation, from the EH action, of the EFE incorporating the cosmological parameter $\Lambda$ purely by geometric nature. After this, we will obtain the equation of SW due to the back reaction of the boundary $\partial V$. Then, we assume that the tetra-vector $\delta W^{\alpha}$ can be written as the contraction of a tensor field, that is: 
\begin{equation}\label{eq:36}
\delta W^{\alpha} = g^{\beta\gamma}\delta\Psi^{\,;\alpha}_{\,\,\,\beta\gamma} \,.
\end{equation}
Thus, the flux $\delta \Phi$ is redefined as:
\begin{equation}\label{eq:40}
\delta W^{\alpha}_{\,\,;\alpha}= \delta \Phi = g^{\beta\gamma}\Box\delta \Psi_{\beta\gamma} \,,
\end{equation}
where $\Box \equiv \nabla_{\mu}\nabla^{\mu}$ is the D'Alembertian. Vibration modes of SW are related to the tensor sector, hence the introduction of the tensor $\delta \Psi_{\beta\gamma}$. Now we can construct a relation between them with respect to the variation of the line element $\delta S$, in the following way:
\begin{equation}\label{waveeq0}
\frac{\delta\Phi}{\delta S} = \Lambda g^{\alpha\beta}\frac{\delta g_{\alpha\beta}}{\delta S} = g^{\alpha\beta}\Box \frac{\delta \Psi_{\alpha\beta}}{\delta S} \,,   
\end{equation}
where we have used $2\delta (\ln\sqrt{-g}) = 2\delta(\sqrt{-g})/\sqrt{-g} = g^{\mu\nu}\delta g_{\mu\nu}$, and here $\delta(\sqrt{-g})=\sqrt{-g} g^{\mu\nu}\delta g_{\mu\nu}/2$. Note that we have assumed that the variation and the D'Alembertian commute with each other. We can simplify eq.~(\ref{waveeq0}), so we define an auxiliary tensor field: 
\begin{equation}\label{chi_definition}
\chi _{\alpha\beta} \equiv \frac{\delta \Psi_{\alpha\beta}}{\delta S} \,,    
\end{equation}
then consider that $g^{\alpha\beta}\Box\chi_{\alpha\beta} = \Box g^{\alpha\beta}\chi_{\alpha\beta} \equiv \Box \chi$, where $\chi$ is the trace of $\chi_{\alpha\beta}$. Finally, we have the following: 
\begin{equation}\label{eq:56}
\Box \chi = \Lambda g^{\alpha\beta}\frac{\delta g_{\alpha\beta}}{\delta S} \,.
\end{equation}
The above expression is the wave equation to solve. The right-hand side describes the source term coming from the perturbations of the geometry times the $\Lambda$ constant, associated with the back-reaction effects. 
\section{Partial collapse of a compact object}\label{toy_model}
In this section we present a toy model of a partial collapse of a compact object and the emission of SW due to back-reaction effects. We propose as a source mass term a non-smooth continuous function that describes a mass-loss. Generally, GW are produced by symmetry breaking, so we consider a spherical non-rotating compact object, which loses mass, here is the rupture of symmetry, as an attempt to relate SW to GW. 

Now, we introduce the framework and setting of our problem. We consider a specific scenario, in which the initial mass is $M_0$, and the final mass is $M_0/2$. We have made this choice to avoid the final singularity from the total collapse of the object before it reaches the Schwarzschild radius. There is a time $t_c$ when the collapse finishes, and the expelled mass from the star decreases exponentially with time. After $t_c$, the mass remains constant and equal to $M_0/2$. Hence, the expression for the mass is:
\begin{equation}\label{eq:57}
M(t) = \left\{
        \begin{array}{ll}
        M_0 e^{-bt} & \quad t \leq t_c \,,\\
        \frac{1}{2}M_0 & \quad t > t_c \,.
        \end{array}
\right.
\end{equation}
Note that this function represents the evolution of the mass and must be continuous at $t_c$, so by introducing this condition we can obtain the value of the collapse time, which is $t_c = (\ln 2)/b$; hence, the parameter $b$ determines how fast or slow the collapse will be. We assume that the geometry right before the collapse is the Schwarzschild metric:
\begin{equation}\label{eq:58}
ds^2 = g_{\mu\nu}dx^{\mu}dx^{\nu} = \beta_{0}(r)dt^2-\frac{dr^2}{\beta_{0}(r)}-r^2d\Omega ^2 \,,
\end{equation}
with $\beta_{0}(r)= 1-2GM_0/r$, and the usual solid angle element $d\Omega ^2 = d\theta^2+\sin^2\theta d\phi^2$. Then, we compute the D'Alembertian: 
\begin{eqnarray}
&& \Box \chi = g^{\alpha\beta}\nabla_{\alpha}\nabla_{\beta}\chi(t,r,\theta,\phi) =  \nonumber \\
&& - \beta_{0}(r)\frac{\partial^{2}}{\partial r^{2}} \chi{\left(t,r,\theta,\phi \right)} - \left(\frac{\beta_{0}(r)+1}{r}\right)\frac{\partial}{\partial r} \chi{\left(t,r,\theta,\phi \right)} + \frac{1}{\beta_{0}(r)}\frac{\partial^{2}}{\partial t^{2}} \chi{\left(t,r,\theta,\phi \right)} \nonumber\\
&& - \frac{1}{r^{2}}\frac{\partial^{2}}{\partial \theta^{2}} \chi{\left(t,r,\theta,\phi \right)} - \frac{1}{r^{2} \tan\theta}\frac{\partial}{\partial \theta} \chi{\left(t,r,\theta,\phi \right)} - \frac{1}{r^{2} \sin^{2}\theta}\frac{\partial^{2}}{\partial \phi^{2}} \chi{\left(t,r,\theta,\phi \right)} \label{dalembertian}\,.
\end{eqnarray}
Above expression (eq.~(\ref{dalembertian})) determines the left hand side of eq.~(\ref{eq:56}). On the other hand, the Schwarzschild metric changes during the transition (collapse), and we then propose that the mass lost of the object is transferred to the SW as energy. In that way, the metric during the collapse becomes:
\begin{equation}\label{eq:59}
ds^2 = \beta(r,t)dt^2-\rho^2(t)\left[\frac{dr^2}{\beta(r,t)}+r^2d\Omega ^2\right] \,.
\end{equation}
Now, the function $\beta(r,t)$ depends on time: $\beta(r,t)=1-2GM(t)/r$. Then $\rho(t) =KGM(t)$ describes how the radius of the object decreases, but recalling that its size stays above the Schwarzschild's radius ($r_{S}=2GM_{0}$), even at the end of the transition, and here $K=\rho_{0}/(GM_{0})>2$, where $\rho_0 = \rho(t_c)$. Thus, we have the right hand side of eq.~(\ref{eq:56}):
\begin{equation}\label{source}
\Lambda g^{\alpha\beta}\frac{\delta g_{\alpha\beta}}{\delta S} = \frac{6\Dot{M}(t)}{M(t)}\sqrt{\frac{r}{r-2GM(t)}}\Lambda \,,   
\end{equation}
where we have used $d/dS = U^{\alpha}d/dx^{\alpha}$, and for a co-moving observer $U^{0}=\sqrt{g^{00}}=1/\sqrt{\beta(r,t)}$, and $U^{i}=0$. Moreover, to simplify mathematically the source term (eq.~(\ref{source})) we take $r = \alpha G M(t)$, where the parameter $\alpha$ quantifies the radius size at sufficiently large times, and $\alpha$ has to be larger than the minimum value of the event horizon $r_{S}$. Therefore, explicitly the equation to solve is:
\begin{eqnarray}
&&  \beta_{0}(r)\frac{\partial^{2}}{\partial r^{2}} \chi{\left(t,r,\theta,\phi \right)} + \left(\frac{\beta_{0}(r)+1}{r}\right)\frac{\partial}{\partial r} \chi{\left(t,r,\theta,\phi \right)} - \frac{1}{\beta_{0}(r)}\frac{\partial^{2}}{\partial t^{2}} \chi{\left(t,r,\theta,\phi \right)} \nonumber\\
&& + \frac{1}{r^{2}}\frac{\partial^{2}}{\partial \theta^{2}} \chi{\left(t,r,\theta,\phi \right)} + \frac{1}{r^{2} \tan\theta}\frac{\partial}{\partial \theta} \chi{\left(t,r,\theta,\phi \right)} + \frac{1}{r^{2} \sin^{2}\theta}\frac{\partial^{2}}{\partial \phi^{2}} \chi{\left(t,r,\theta,\phi \right)} =  6b\Lambda \sqrt{\frac{\alpha}{\alpha-2}} \label{full_equation}\,.
\end{eqnarray}
In the next segment we will solve eq.~(\ref{full_equation}) by the method of separation of variables. 
\subsection{Solution of the wave equation}
The problem setting is already established from eq.~(\ref{dalembertian}) (before the collapse) and eq.~(\ref{source}) (during the transition). Having all these ingredients, the complete solution for SW can be written in the form of series for the field $\chi$~\cite{HERNANDEZ2020100424}:
\begin{equation}\label{eq:60}
\chi (t,r,\theta,\phi) = \sum_{n=0}^{\infty}\chi_{n} (t,r,\theta,\phi) \,,
\end{equation}
and to obtain the modes we take the separation of variables method, that is~{\footnote{The full solution consists of two functions, the homogeneous and particular solutions: $\chi_{nlm} (t,r,\theta,\phi) = \chi^{H}_{nlm} (t,r,\theta,\phi) + \chi^{P}_{nl} (t,r,)$, where the particular one only depends on $r,t$.}}:
\begin{equation}\label{eq:62}
\chi_{nlm} (t,r,\theta,\phi) = R_{nl}(r,t)Y_{lm}(\theta,\phi) \,,
\end{equation}
so we obtain the differential equations: 
\begin{eqnarray}
&& -\frac{r^{2}}{\beta_{0}}\frac{\partial^{2} R_{nl}(t,r)}{\partial t^{2}} + r^{2}\beta_{0}\frac{\partial^{2} R_{nl}(t,r)}{\partial r^{2}} + r(\beta_{0}+1)\frac{\partial R_{nl}(t,r)}{\partial r} + l(l+1)R_{nl}(t,r) = 6b\Lambda r^{2} \sqrt{\frac{\alpha}{\alpha-2}} \,, \label{eqRnl} \\ 
&& \frac{1}{\sin\theta}\frac{\partial}{\partial\theta}\left(\sin\theta\frac{\partial Y_{lm}(\theta,\phi)}{\partial\theta} \right) + \frac{1}{\sin^{2}\theta}\frac{\partial^{2} Y_{lm}(\theta,\phi)}{\partial\phi^{2}} + l(l+1)Y_{lm}(\theta,\phi) =0 \label{eqYlm}\,,   
\end{eqnarray}
where the functions $Y_{lm}(\theta,\phi)$ are the spherical harmonics:
\begin{equation}\label{eq:63}
Y_{lm}(\theta,\phi) = (-1)^{m}\sqrt{\frac{(2l+1)(l-m)!}{4\pi (l+m)!}}P^{m}_{l}(\cos\theta)e^{im\phi} \,,
\end{equation}
which are the solutions of the angular part of eq.~(\ref{dalembertian}); here $!$ denotes the factorial and $P^{m}_{l}(\cos\theta)$ are the Legendre polynomials. Hence, the solution of eq.~(\ref{eqRnl}) is:
\begin{eqnarray}
&& \hspace{-1.5cm} R_{nl}(t,r)  =  c_{1}\sqrt{r}\left[r^{-\frac{\sqrt{-4 l^{2}-4 l +1}}{2}} \,_{2}F_{1}^{(1)} + r^{\frac{\sqrt{-4 l^{2}-4 l +1}}{2}} \,_{2}F_{1}^{(2)}\right] \nonumber \\ 
&& + c_{2}e^{\left(r - t \right) \eta}\left[\left(2 G M_{0} -r \right)^{2 G M_{0} \eta} H_{C}^{(1)} + \left(2 G M_{0} -r \right)^{-2 G M_{0}\eta} H_{C}^{(2)} \right] \nonumber \\
&& + \frac{96 \Lambda  b}{\left(l^{2}+l+6\right)} \left[\frac{r^{2}}{16}+\frac{ G M_{0} r}{2\left(l^{2}+l+2\right) } + \frac{G^{2} M_{0}^{2}}{l \left(l +1\right) \left(l^{2}+l+2\right) }\right] \sqrt{\frac{\alpha}{\alpha -2}} \,, \label{solRnl}
\end{eqnarray}
where $\eta$ is the constant of the method of separating variables, and $c_{1,2}$ are integration constants. Then, the expressions $\,_{2}F_{1}^{(1,2)}$ are the hypergeometric and $H_{C}^{(1,2)}$ the confluent Heun functions, respectively, and are given by: 
\begin{eqnarray}
&& \,_{2}F_{1}^{(1,2)} = \,_{2}F_{1} \left[\frac{1}{2} \pm \frac{\sqrt{-4 l^{2}-4 l +1}}{2},\frac{1}{2} \pm \frac{\sqrt{-4 l^{2}-4 l +1}}{2};1\pm\sqrt{-4 l^{2}-4 l +1},\frac{2 G M_{0}}{r}\right] \,,  \\
&& H_{C}^{(1,2)} =  H_{C} \left(-4 G M_{0} \eta,\pm 4 G M_{0} \eta,0,8 \eta^{2} G^{2} M_{0}^{2},-8 \eta^{2} G^{2} M_{0}^{2}+l^{2}+l ,\frac{2 G M_{0} -r}{2 G M_{0}}\right) \,.
\end{eqnarray}
One expects a decay behaviour of the function $|R_{nl}(t,r)|$ at $t\gg 1$~\cite{HERNANDEZ2020100424}, therefore, the only relevant solution must have the $-\eta t$ term in the exponential function. Furthermore, the $\eta$ parameter must be complex; otherwise, the Heun functions are indeterminate. 

\begin{figure}[htbp] 
\includegraphics[scale=0.75]{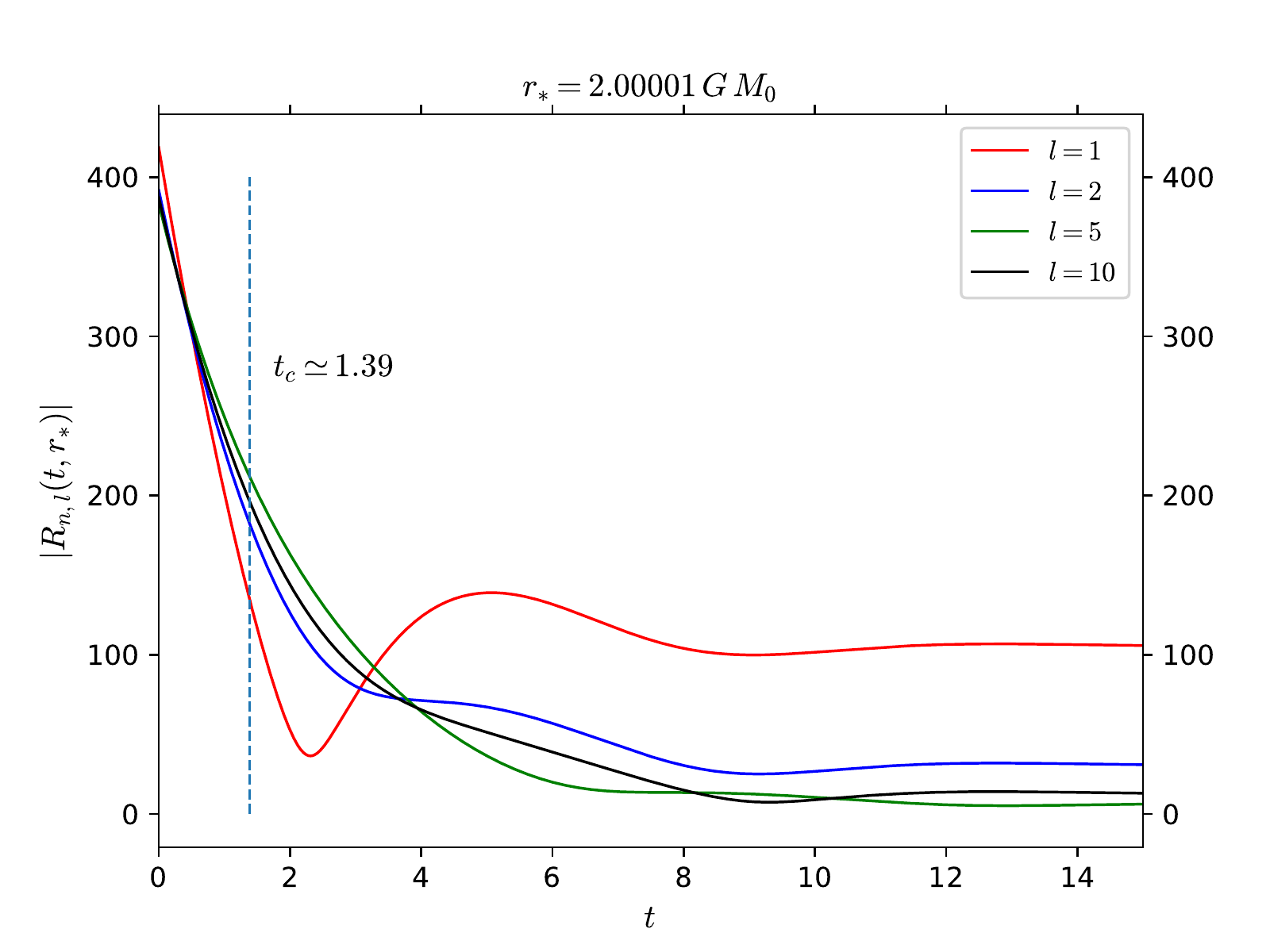}
\caption{Evolution of the norm $|R_{nl}(t,r_{*})|$ in terms of $t$. We fix the values $c_{1}=c_{2}=1$, $\eta=\sqrt{-0.5-0.75i}$, and $\alpha = 2.5$. We also take $\Lambda=3/b^{2}$, with $b=0.5$; and we place ourselves outside the compact object at a distance $r_{*}=2.00001 GM_{0}$, which is, in fact, very close to $r_{S}$, where $G=1$ and $M_{0}=M_{\odot}=1$. We show the result for different values of $l=$ 1 (red), 2 (blue), 5 (green), and 10 (black). Note that time is measured in reduced Planck units since $8\pi G = 1$.}\label{fig:Rnl_0alpha}
\end{figure}

\begin{figure}[htbp] 
\includegraphics[scale=0.75]{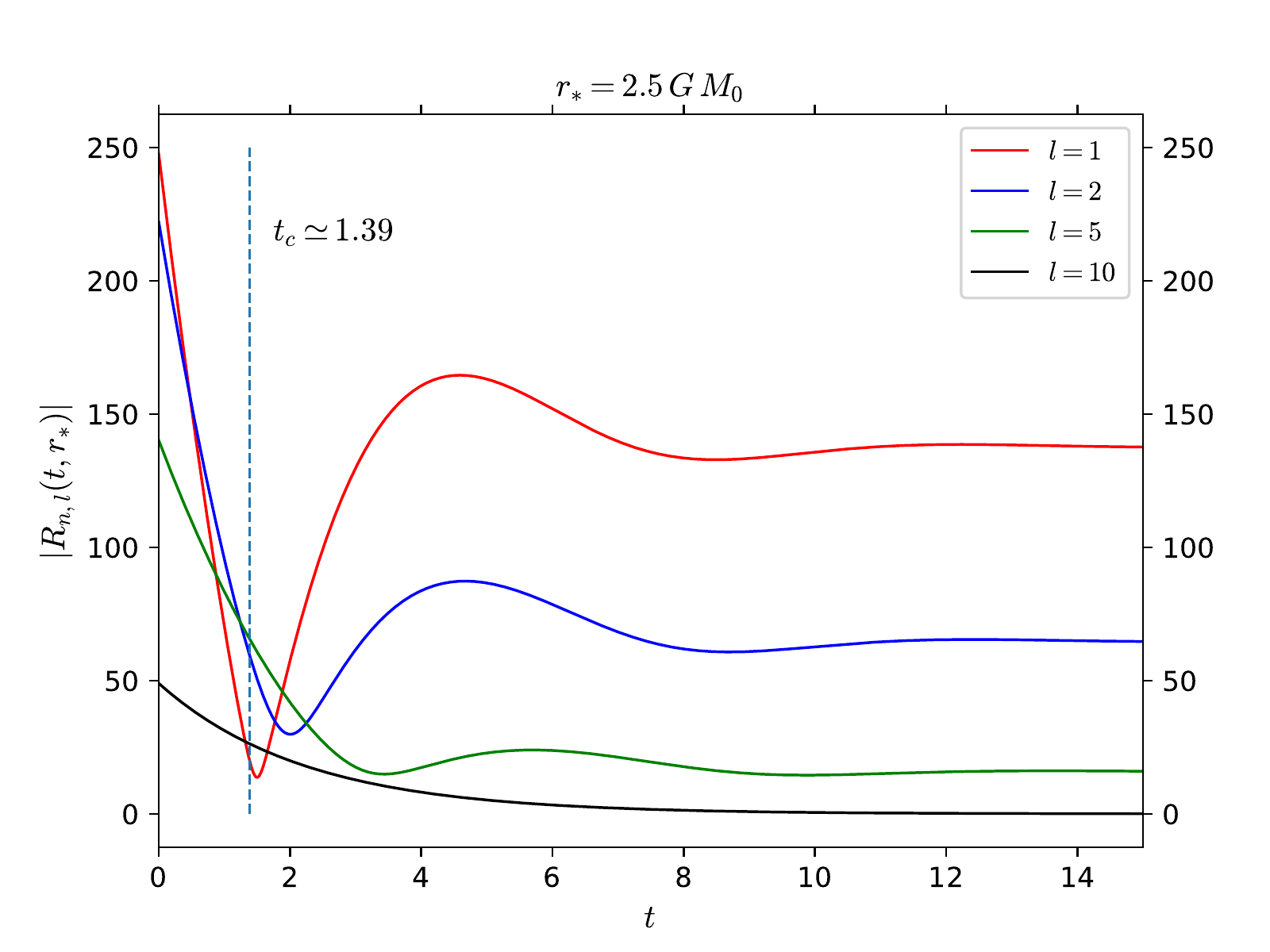}
\caption{Evolution of the norm $|R_{nl}(t,r_{*})|$ in terms of $t$. We fix the values $c_{1}=c_{2}=1$, $\eta=\sqrt{-0.5-0.75i}$, and $\alpha = 2.5$. We also take $\Lambda=3/b^{2}$, with $b=0.5$; and we place ourselves outside the compact object at a distance $r_{*}=2.5 GM_{0}$, which is equal to $\alpha$, where $G=1$ and $M_{0}=M_{\odot}=1$. We show the result for different values of $l=$ 1 (red), 2 (blue), 5 (green), and 10 (black). Note that time is measured in reduced Planck units since $8\pi G = 1$.}\label{fig:Rnl_alpha}
\end{figure}

\begin{figure}[htbp] 
\includegraphics[scale=0.75]{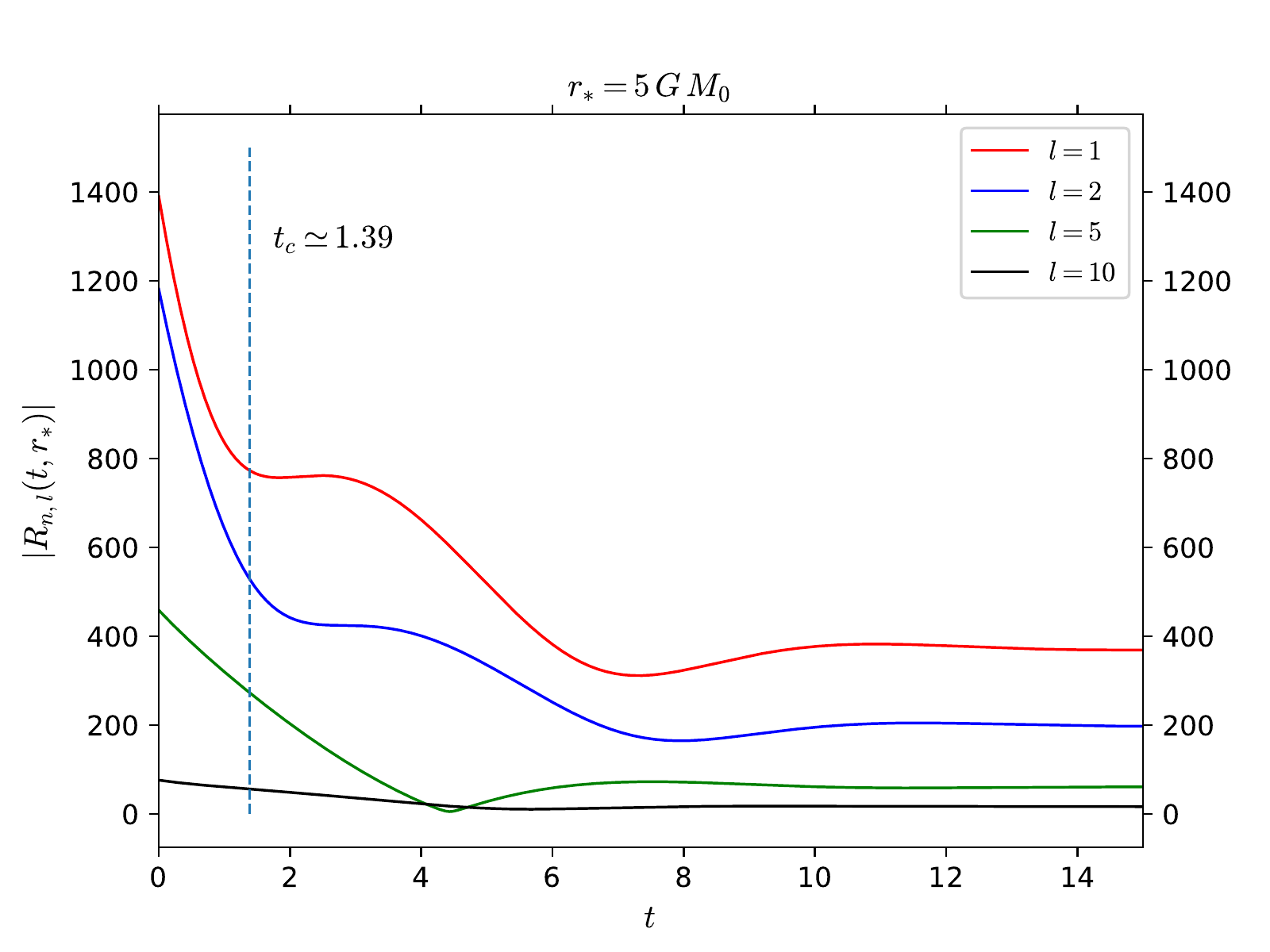}
\caption{Evolution of the norm $|R_{nl}(t,r_{*})|$ in terms of $t$. We fix the values $c_{1}=c_{2}=1$, $\eta=\sqrt{-0.5-0.75i}$, and $\alpha = 2.5$. We also take $\Lambda=3/b^{2}$, with $b=0.5$; and we place ourselves outside the compact object at a distance $r_{*}=5 GM_{0}$, which is equal to $2\alpha$, where $G=1$ and $M_{0}=M_{\odot}=1$. We show the result for different values of $l=$ 1 (red), 2 (blue), 5 (green), and 10 (black). Note that time is measured in reduced Planck units since $8\pi G = 1$.}\label{fig:Rnl_2alpha}
\end{figure}

We present three distinct examples of the evolution of the norm $|R_{nl}(t,r_{*})|$ in terms of $t$: figs.~\ref{fig:Rnl_0alpha}, \ref{fig:Rnl_alpha}, \ref{fig:Rnl_2alpha}. With values $c_{1}=c_{2}=1$, $\eta=\sqrt{-0.5-0.75i}$, and $\alpha = 2.5$. We also take $\Lambda=3/b^{2}$, with $b=0.5$; and we place ourselves outside the compact object at a distance $r_{*}/(G M_{0})=2.00001 \,, 2.5\,, 5$; which are very close to $r_{S}$, equal to $\alpha$, and $2\alpha$, respectively; here $G=1$ and $M_{0}=M_{\odot}=1$. And four different results are shown for the values $l=$ 1 (red), 2 (blue), 5 (green), and 10 (black). In fig.~\ref{fig:Rnl_0alpha} an observer is placed very close to $r_{S}$; and, in fact, one can observe that all signals, regardless the number $l$, began at $t=0$ very close to each other; however, from about $t_{c}$ they evolved differently. Then from figs.~\ref{fig:Rnl_alpha}, \ref{fig:Rnl_2alpha}, can be noticed that the farther away we position ourselves $|R_{nl}(t,r_{*})|$ oscillates less, and, in fact, it only decays as expected. Note that in all figures, the decay behaviour is actually present at $t\gg 1$ (after $t_{c}$), and becomes more evident for larger $l$. In addition, for the results that have smaller $l$'s their amplitudes are larger when the asymptotic character of $|R_{nl}(t,r_{*})|$ clearly appears. Finally, the farther away an observer is set, the fewer oscillations are perceived; however, from our particular fixed set of parameters, the best spot to observe the wiggles of the emitted SW is close to $r_{*}\simeq \alpha$. 

%

\section{Final discussions}
\label{conclusions}

We presented a non-perturbative mechanism of novel production of SW. The Lagrangian formulation of GR, where a manifold holds a boundary $\partial V$ which generates back-reaction effects, yields the generation of SW. Within this framework, we study a toy model of a partial collapse of a compact object and the emission of SW. We propose as a source mass term a non-smooth continuous function that describes a mass-loss.

To first solve our model, we consider a specific scenario in which the initial mass is $M_0$ and the final mass is $M_0/2$. Then at a time $t_c$ the collapse stops and immediately afterward the object reduces half its mass. Also, we assume that the geometry right before the collapse is the Schwarzschild metric; and this geometry changes with time during the transition (collapse), and we then propose that the mass lost of the object is transferred to the SW as energy. Moreover, we simplify even further by taking $r = \alpha G M(t)$ in the source term, where the parameter $\alpha$ quantifies the size of the radius at sufficiently large times, and it has to be larger than the Schwarzschild's radius $r_{S}=2GM_{0}$. 

Mathematically we solve eq.~(\ref{full_equation}) using the method of separation of variables: $\chi_{nlm} (t,r,\theta,\phi) = R_{nl}(r,t)Y_{lm}(\theta,\phi) $. Here, $Y_{lm}(\theta,\phi)$ are the spherical harmonics, and $R_{nl}(r,t)$ allows one to compute the amplitude of the wave: $|R_{nl}(r,t)|$. Having the full solution, three distinct examples of the evolution of the norm $|R_{nl}(t,r_{*})|$ in terms of $t$: figs.~\ref{fig:Rnl_0alpha}, \ref{fig:Rnl_alpha}, \ref{fig:Rnl_2alpha}, are shown. We place ourselves outside the compact object at a distance $r_{*}/(G M_{0})=2.00001 \,, 2.5\,, 5$; which are very close to $r_{S}$, equal to $\alpha$, and $2\alpha$, respectively. And four different results are shown for the parameter $l=1,2,5,10$. Note that in all instances the damping behaviour is actually present after $t_{c}$. In addition, for the results that have smaller $l$'s their amplitudes are larger when the asymptotic character of $|R_{nl}(t,r_{*})|$ clearly appears. Finally, the farther away an observer is set, the fewer oscillations are perceived; however, from our particular fixed set of parameters, the best spot to observe the wiggles of the emitted SW is close to $r_{*}\simeq \alpha$.  

In the future, a very good exercise would be to check whether this formalism in fact matches to the results, already established, of GW production from the linearised theory of GR and, hence, this new scheme gains a solid theoretical ground.

\acknowledgments
This work was supported by CONACyT Network Project No. 376127 {\it Sombras, lentes y ondas gravitatorias generadas por objetos compactos astrofísicos}. R.H.J is supported by CONACYT Estancias Posdoctorales por M\'{e}xico, Modalidad 1: Estancia Posdoctoral Acad\'{e}mica and by SNI-CONACYT. C.M. thanks PROSNI-UDG support. 
%
%
\bibliography{esw}

\end{document}